# STABLE DRUG DESIGNING BY MINIMIZING DRUG PROTEIN INTERACTION ENERGY USING PSO

Anupam Ghosh[1] Mainak Talukdar[2] and Uttam Kumar Roy[3]

[1]Indian Institute of Technology Bombay, Mumbai, India
`anupam.ghsh@gmail.com`
[2]Cognizant Technology Solutions, Kolkata, India
`codes.mnk@gmail.com`
[3]Jadavpur University, Kolkata, India
`royuttam@gmail.com`

## ABSTRACT

*Each and every biological function in living organism happens as a result of protein-protein interactions. The diseases are no exception to this. Identifying one or more proteins for a particular disease and then designing a suitable chemical compound (known as drug) to destroy these proteins has been an interesting topic of research in bio-informatics. In previous methods, drugs were designed using only seven chemical components and were represented as a fixed-length tree. But in reality, a drug contains many chemical groups collectively known as pharmacophore. Moreover, the chemical length of the drug cannot be determined before designing the drug.*

*In the present work, a Particle Swarm Optimization (PSO) based methodology has been proposed to find out a suitable drug for a particular disease so that the drug-protein interaction becomes stable. In the proposed algorithm, the drug is represented as a variable length tree and essential functional groups are arranged in different positions of that drug. Finally, the structure of the drug is obtained and its docking energy is minimized simultaneously. Also, the orientation of chemical groups in the drug is tested so that it can bind to a particular active site of a target protein and the drug fits well inside the active site of target protein. Here, several inter-molecular forces have been considered for accuracy of the docking energy. Results show that PSO performs better than the earlier methods.*

## 1. INTRODUCTION

Protein is a macro molecule primarily consisting of amino acids [1]. Protein is highly responsible for structural and functional characteristics of cells, and communication of biological signals among cells. All proteins available in the nature are not useful for the living organism. However, every biological function in living organism happens as a result of protein-protein interactions [2, 3]. There are evidences of proteins which cause fatal or infective diseases. Researchers take keen interest to identify appropriate drug that fits well in the active sites of harmful protein. These drugs, which can bind in the active site of target protein and therefore change the functional behavior of that particular protein, are called *ligands*. This mechanism of binding of drug with the target protein is called *docking* [4]. The challenge is to predict an accurate structure of the ligand when the active site configuration of the target protein is known.

Docking is very much important in cellular biology. In docking, proteins interact with themselves and with other molecular components. It is the key concept to oriental drug design. The result of docking can be used to find inhibitors for specific target proteins and thus to design new stable drugs. Protein-ligand docking is an energy minimization search and optimization problem with the aim to find the best ligand conformation for active sites of a target protein.

In this paper, the scope of the well-known Particle Swarm Optimization (PSO) algorithm [5, 2, 6, 7] is studied to determine the ligand structure to be docked at the active site of the target protein. The choice of PSO in the present context is inspired by social behavior of bird flocking and fish schooling [8]. It begins with a random population and then searches for optimal solution by updating the population based on fitness of the current solution.

Evolutionary computation is used to place functional groups in the nodes of the variable tree structured ligand. The tree structure helps in connecting primitive fragments or radicals to determine the right candidate solution for the ligand that best suits with the active site of the protein. The PSO based algorithm flows through the entire search space, and records the best individuals they have met. At each step, it changes its position according to the best individuals to reach a new position. In this way, the whole population evolves towards the optimum, step by step.

A 'fitness function' is generally introduced in a meta-heuristics algorithm to determine the quality of the desired solutions for an optimization problem. Naturally, the better the formulation of the fitness function, the better is the expected quality of the trial solutions. In the present context of the ligand-docking problem, optimal selection of the ligand is inspired by minimization of an energy function that determines the stable connectivity between the protein and the ligand. So, the fitness function here is an energy function, whose minimization yields trial solutions to the problem. One important factor in the field of drug design is the identification of proper ligand. In general, one or more proteins are typically involved in the bio-chemical pathway of a disease. The treatment aims to appropriately identify those proteins and reduce their effects by designing a ligand molecule [9] that can bind to protein's active site. That is, the structure of a ligand molecule is evolved from a set of groups in close proximity to crucial residues of the protein; a molecule is thereby designed that fits the protein target receptor such that a criterion for Van Der Waals & electrostatic interaction energy is optimized. We propose to adopt this approach in this paper.

In this paper, we propose a novel approach by which the drug is represented by a variable length tree-like structure, forty-five functional groups, Van Der Waals as well as electrostatics force to design the ligand structure. We have applied our approach to Human Rhinovirus stain 14, Plasmodium Falciparum and HIV-1 protease viruses. We have significantly improved the work done by earlier researchers due to following reasons 1) A new representation for the ligand is used; 2) Both Van Der Waal and electrostatic energies are optimized and 3) PSO is used because of its ability to handle to the ligand design problem. It can be seen that PSO method has performed quite satisfactorily in comparison with Genetic Algorithm (GA) (fixed length tree) [10], Neighbourhood Based Genetic Algorithm (NBGA) (fixed length tree) [11], and variable length Neighbourhood Based Genetic Algorithm (VNBGA) (variable length tree) [12].

The rest of the paper is organized as follows. Section 2 briefs the related work. Section 3 contains the formulation of protein–ligand docking problem; section 4 depicts the principles used to predict the ligand structures. In section 5, PSO algorithm used to find the best ligand structure is described. The pseudo-code for solving the given constrained optimization function is given in section 6. A comparison of experiment results for 3 proteins in section 7 is also given. Section 8 concludes the paper with the future work.

## 2. RELATED WORK

A fixed length Genetic Algorithm approach for ligand design was used in [10, 8, 13, 14] to evolve molecular structure of possible ligand that binds to a given target protein. The drug is represented by a fixed tree-like structure, comprising of molecules at the nodes and the bonds as links. Evidently, an a priori knowledge of the size and length of the tree is difficult to obtain before the experiment.

Another approach for ligand design, which is based on variable length representation of trees on both sides of the pharmacophore was studied by Bandopadhyay et al [12]. However, the approach is restricted to build the ligand in two-dimensional space from a small suite of seven functional groups. Furthermore, the fitness value of the ligand is confined to Van Der Waals force only.

One more approach for ligand design, that is based on the presence of a fixed pharmacophore and that uses the search capabilities of genetic algorithms, was studied by Goah and Foster [10], where the harmful protein human Rhinovirus strain14 was used as the target. This pioneering work assumed a fixed tree structure representation of the molecule on both sides of the pharmacophore. Evidently, an a priori knowledge of the size of the tree is difficult to obtain. Moreover, it is known that no unique ligand structure is best for a given active site geometry.

Furthermore, few more algorithms were designed by researchers for ligand design *viz*. AutoDock 3.05, SODOCK and PSO@AUTODOCK. In [6], a novel optimization algorithm SODOCK is used for solving protein–ligand docking problems. And results are compared with the performance of AutoDock 3.05 [15]. SODOCK is based on particle swarm optimization and a local search strategy. The work uses the environment and energy function of AutoDock 3.05. Results show that SODOCK performs better than AutoDock in terms of convergence performance, robustness, and obtained energy.

In [16], a novel meta-heuristic algorithm called PSO@AUTODOCK based on varCPSO and varCPSO-ls is used to calculate the docking energy. varCPSO stands for velocity adaptive and regenerative CPSO. varCPSO-ls is varCPSO with local search technique to avoid convergence at local minima. In addition to that, comparison is done with other docking techniques as GOLD 3.0, DOCK 6.0, FLEXX 2.2.0, AUTODOCK 3.05, and SODOCK. It has been shown that PSO@AUTODOCK gives highly efficient docking program in terms of speed and quality for flexible peptide–protein docking and virtual screening studies.

In our paper, we propose a PSO based approach wherein a variable length tree-like structure, forty-five functional groups, Van Der Waals and electrostatics force is used to design the ligand structure. We have compared our PSO based approach with GA (fixed length tree), NBGA (fixed length tree) and VNBGA (variable length tree).

## 3. FORMULATION OF THE PROBLEM

In protein-ligand docking problem, the objective is to minimize the energy. Firstly, the internal energy of the ligand should be minimized for better stability of the ligand [17]. The inter-molecular energy value, which is thereafter optimized, is the interaction energy between the ligand and the active site of the receptor protein. This energy calculation is based on the proximity of the different residues in the active site of the receptor protein to the closest functional groups in the ligand and their chemical properties. The inter-molecular interaction energy is computed in terms of the Van Der Waals energy and the electrostatic energy.

It is noted that the distance between a residue of the target protein receptor and its closest functional group should be between $0.7 \overset{0}{A}$ and $2.7 \overset{0}{A}$. If the functional group and closest residue are of different polarity, then electrostatic force of attraction also acts between those molecules. The interaction energy of the ligand with the protein is the sum of Van Der Waals Force and electrostatic energy [4].

$$E_{electrostatics} = \frac{q_A q_B}{4\pi\varepsilon_0 r_{AB}} \quad (1)$$

Value of $\frac{1}{4\pi\varepsilon_0} = 9*10^9 \, N-m^2/c^2$

Where $q_A$ and $q_B$ are the charges of the two atoms is the separation, $\varepsilon$ is the dielectric constant of the surrounding medium $r$ is the distance between charges.

$$E_{vdw} = \frac{A}{r^{12}} - \frac{B}{r^6} \quad (2)$$

Where *A* and *B* are constants and $r$ is distance between the molecules. Value of *A* and *B* depends on the atom pair.

$$E_{total} = E_{electrostatics} + E_{vdw} \quad (3)$$

The fitness value is taken as $F = \frac{1}{E_{total}}$

## 4. FORMATION OF A LIGAND

In the proposed work, we consider that the ligands are built using the fragments from the suite as mentioned in Figure 4. Proteins with known active site configuration are used for evolving ligand structures. The specific target is the known antiviral binding site of the Human Rhinovirus strain 14, Plasmodium Falciparum and HIV-1 Protease. The active site of Human Rhinovirus strain 14 is known as the VP1 barrel as it resembles with a barrel. The molecules which can easily be fit in the structure having minimum interaction energy will be the evolved drug (i.e. ligand). For simplification; a 2-dimensional structure is chosen.

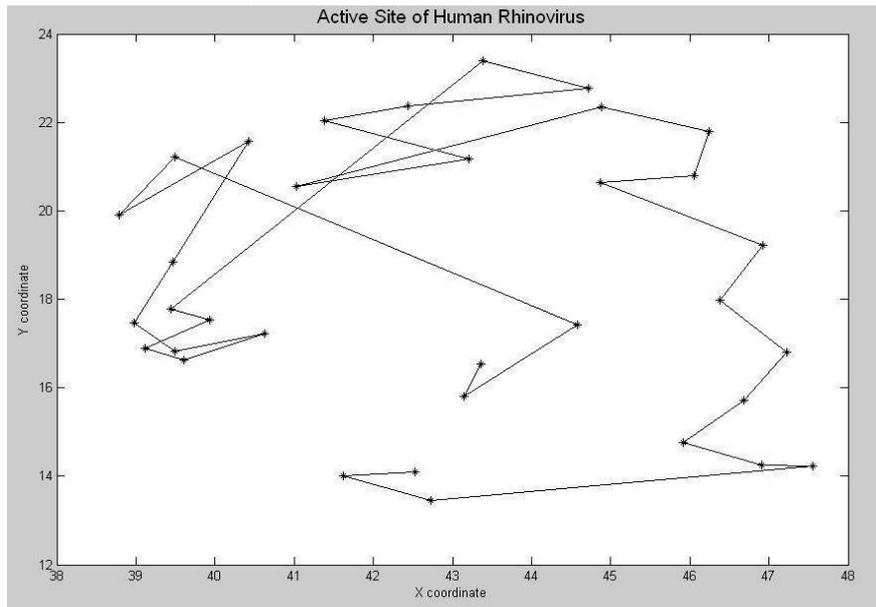

Figure 1. Active site of Human Rhinovirus stain 14

Figure 1 illustrates the active site of Human Rhinovirus strain 14, Figure 2 illustrates the active site of Plasmodium Falciparum and Figure 3 illustrates the active site of HIV-1 protease. For designing the ligand, the co-ordinates of the active site of the protein must be known.

In the present context, we represent a ligand as a variable length tree structure. The length of the ligand will be determined according to the active site of the target protein as it cannot be determined before finding its structure; this can be seen in Figure 1, Figure 2 and Figure 3. We have considered the variable length tree having total 15 nodes. These nodes will be filled by at most 15 functional groups appropriately selected from the set of forty-five functional groups given in Figure 4.

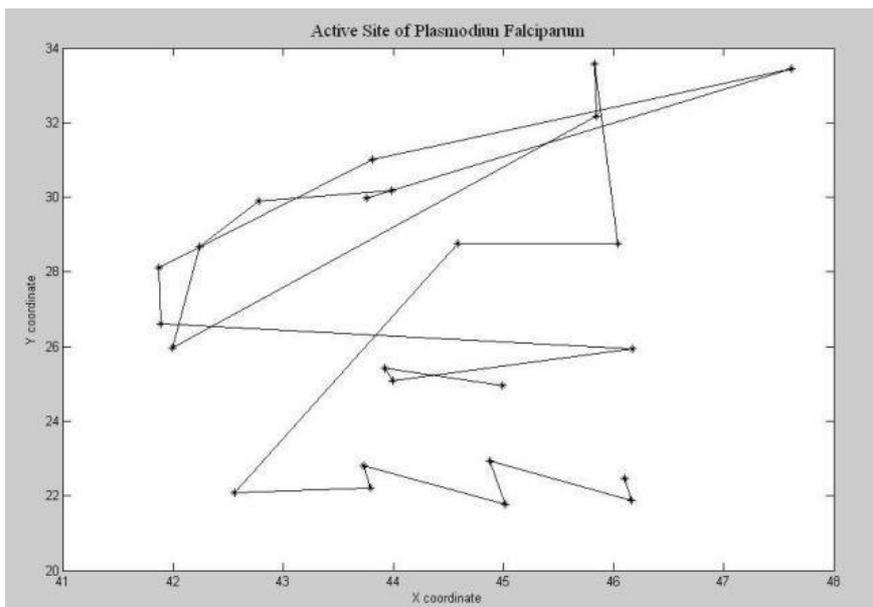

Figure 2. Active site of Plasmodium Falciparum

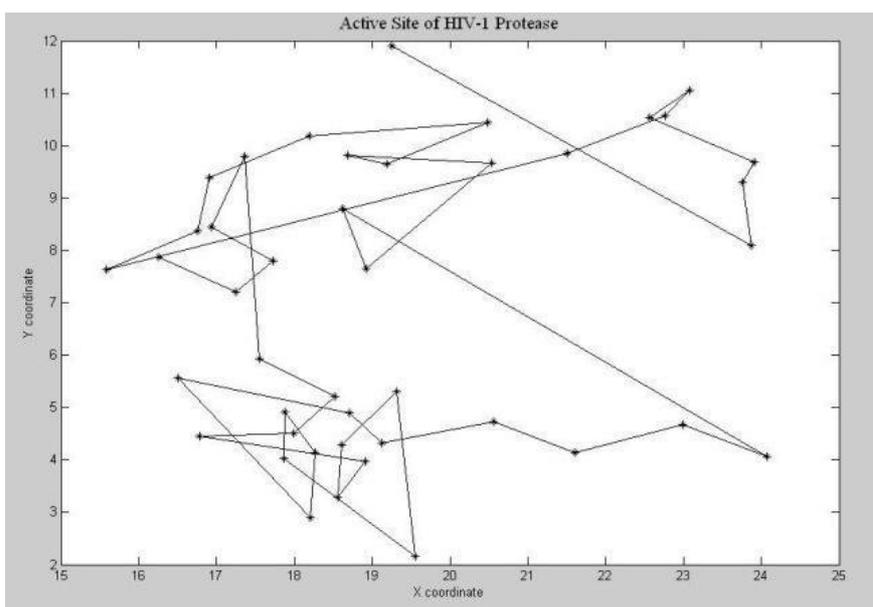

Figure 3. Active site of HIV-I protease

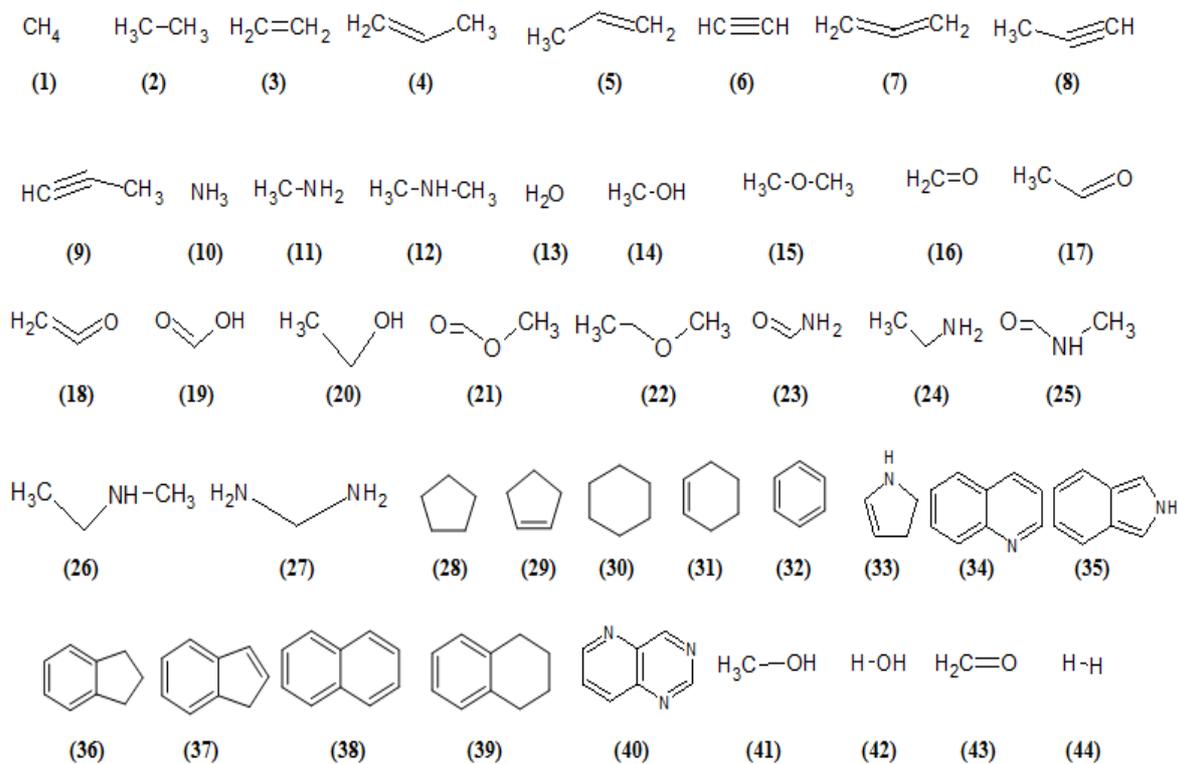

Figure 4. A set of forty five groups used to build ligands (45[th] group is the NULL group)

## 5. PARTICLE SWARM OPTIMIZATION ALGORITHM

PSO or Particle Swarm Optimization algorithm is an evolutionary computation technique that was developed by DT Eberhart and Dr. Kennedy [18]. This algorithm is inspired by social behavior of bird flocking and fish schooling. It also begins with a random population and searches for optima by updating the population. In PSO, the potential solutions, called particles, flow through the whole search space by following the current optimum particles. Each particle has a speed vector and position vector to represent a possible solution.

PSO uses a simple rule. Before a change, each particle has three choices: (1) insists on oneself (2) moves towards the optimum it has met (3) moves towards the best the population has met. PSO reaches to a balanced position among these three choices. The algorithm is described as follows:

$$V_{i,d}^{k+1} = W \times V_{i,d}^{k} + C_1 \times R_1 \times (P_{i,d}^{pBest} - P_{i,d}^{k}) + C_2 \times R_2 \times (P_{d}^{gbest} - P_{i,d}^{k}) \qquad (4)$$

$$P_{i,d}^{k+1} = P_{i,d}^{k} + V_{i,d}^{k+1} \qquad (5)$$

Where $V$ is the velocity, $P$ is the position, $k$ is number of iterations, d is number of particles, $i$ is number of particle dimensions, $W$ is inertia factor. $C_1$ and $C_2$ are acceleration coefficients. $R_1$ and $R_2$ are two other constants in the interval (0, 1).

## 6. SOLVING THE CONSTRAINT OPTIMIZATION PROBLEM USING PSO

Input:
  1. Coordinates of active site.
  2. Valency and length of 45 functional groups.

Output: Desired ligand structure L for receptor target protein P.

*Begin*
    Call PSO (active_site_P);
*End*

Procedure PSO (active_site_P)
*Begin*
  1. Initialize each particle Xi with 15 functional groups randomly chosen from the set of 45 groups at Figure 4.
  2. Assign initial coordinates to each particle and call calculate energy ($X_i$) to find respective fitness.
  3. Calculate fitness of each particle.
  4. Calculate velocity and position of PSO parameters.
  5. Update pBest and gBest as necessary.
  6. Repeat steps 2 to 5 until a convergence criterion is satisfies or maximum iteration ends.
*End*

## 7. EXPERIMENTS AND RESULTS

The experiment was carried out in a simulated environment using MATLAB 2011. Population size for PSO is taken as 50 and the algorithm is run for 100 generation. In each generation, each of the particles is decoded to obtain the corresponding drug structure. Results are taken for different possible positions of the drug within the active site, and the evolved drug having the lowest energy value is taken as the solution. The two dimensional structure of the ligand is drawn using ChemSketch software [19].

We have applied our proposed PSO based approach on three different proteins viz. Rhino virus 14 Mutant N1105S (1RUC), Plasmodium Falciparum (TS-DHFR) (3DGA) and HIV-1 Protease (1W5X). Information about these three proteins is obtained from Protein Data Bank (http://www.rcsb.org/pdb/home/home.do) and the active site of a protein is obtained from http://dogsite.zbh.uni-hamburg.de/ . Table 1 shows the diseases caused by each of these proteins.

Table 1. Protein and corresponding disease caused by the protein.

| Name of the protein | Disease caused by the protein |
|---|---|
| Rhino virus 14 Mutant N1105S (1RUC) | Cough and Cold |
| Plasmodium Falciparum (TS-DHFR) (3DGA) | Malaria |
| HIV-1 Protease (1W5X) | AIDS |

We have compared our PSO based approach using Van Der Waals force with other available approaches. The docking energy values of the ligand–protein complexes are calculated for Human Rhinovirus and are given in Table 2. Lower inter-molecular energy values of the ligand imply better stability of the ligand. As seen in Table 2, our PSO based approach provides more stable ligands that are associated with lower energy values. Moreover, the length of the ligand is also checked with the size of active sites. The length of the ligand should always be less than the length of the active site. In our experiments, it was found that the length of the active site was $8.76 \overset{0}{A}$ for Human Rhinovirus 14 Mutant N1105S, while the length of the ligand found using PSO was $8.72 \overset{0}{A}$ which is less than the length of the active site. Hence, the ligand can fit quite well inside corresponding active site.

Table 2. Comparison of Result for Human Rhinovirus

| Serial Number | Algorithm | Docking Energy (Kcal/mol) |
|---|---|---|
| 1. | Fixed length GA | 11.6454 |
| 2. | NBGA (fixed length) | 11.5748 |
| 3. | VNBGA (Variable length) | 8.1046 |
| 4. | Our Proposed Method (Variable length PSO) | 6.5385 |

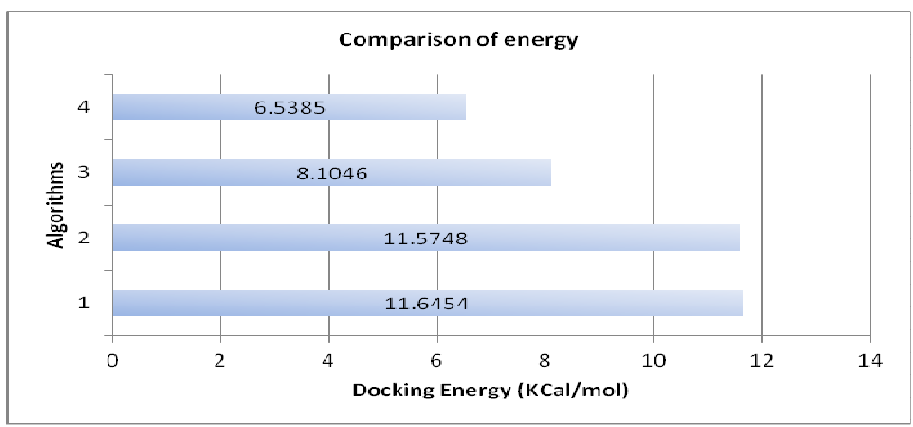

Figure 5. Comparison among algorithms mentioned in Table 2.

We have applied Electrostatics force of attraction along with Van Der Waal force to the Human Rhinovirus; the results obtained are 10.6352 Kcal/mol. However, Van Der Waal force is applied to Plasmodium Falciparum and HIV-1 Protease and the results obtained are 2.5477 Kcal/mol and 13.4303 Kcal/mol respectively. The application of Electrostatics force along with Van Der Waals force for Plasmodium Falciparum and HIV-1 Protease will be kept for future work.

The two dimensional structure of ligand molecule evolved using PSO are pictorially represented in Figure 6, 7 & 8 for Human Rhinovirus strain 14, Plasmodium Falciparum and HIV-1 Protease respectively. It is clear from the figures that the design molecules using PSO fill up the active sites quite well.

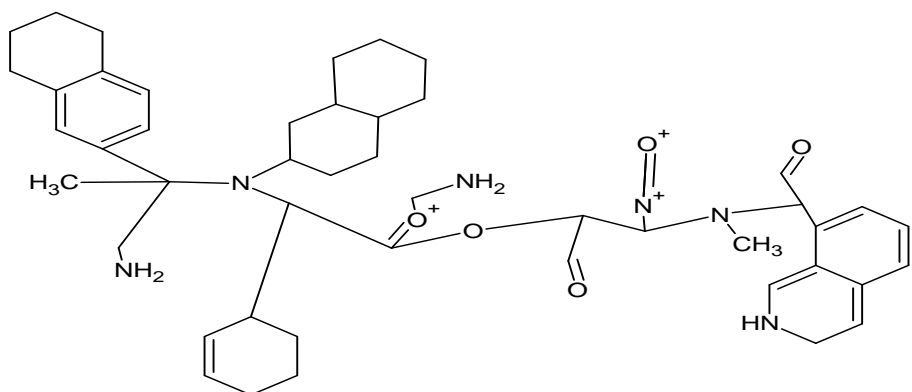

Figure 6. Generated Ligand Structure for Human Rhino Virus

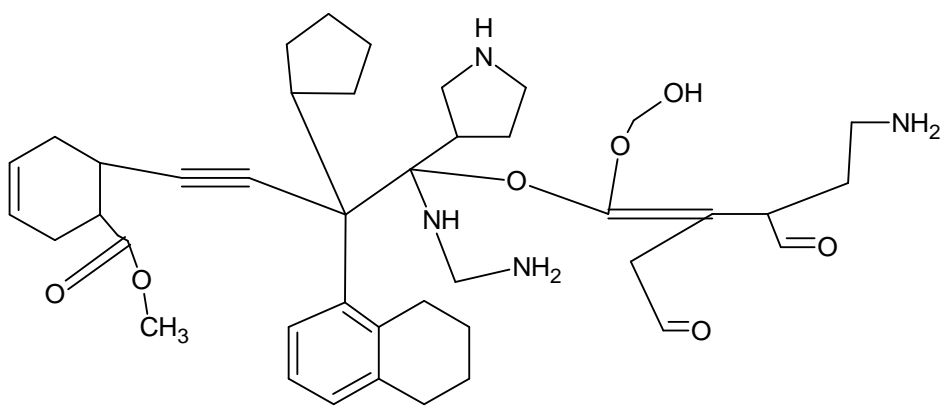

Figure 7. Generated Ligand Structure for Plasmodium Falciparum

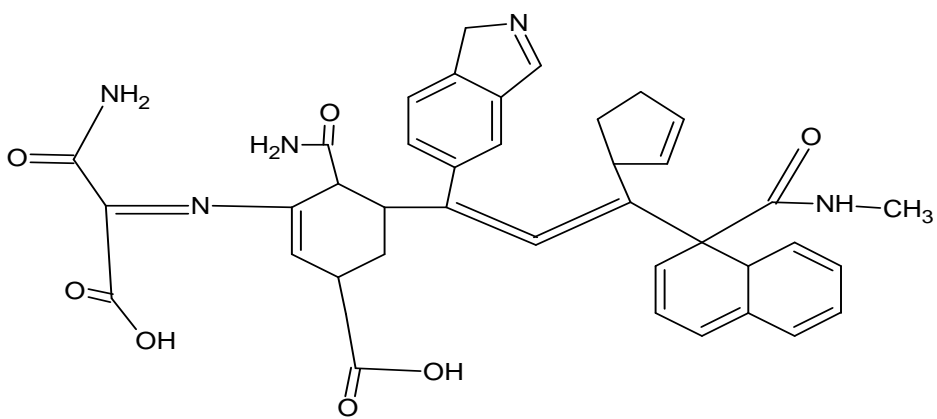

Figure 8. Generated Ligand Structure for HIV-I Protease

## 8. CONCLUSION AND FUTURE WORK

We can conclude that the proposed PSO based algorithm primarily optimizes protein-ligand inter-molecular docking energy. Our algorithm finds the ligand structure in such a way that each ligand can fit into corresponding active site very well. It also gives better result (less energy) than other methods. Moreover, using PSO, we can obtain more stable structure of ligand molecule. This proposed technique can be used to provide a powerful tool for the chemist to evolve molecular structure of ligand once the functional protein is given.

This work uses a two dimensional approach which has some limitations. A three dimensional approach using new data structure for the ligand will be our future research goal. Moreover, we have considered only inter-molecular forces but there are a number of intra-molecular forces like Bond stretching, angle binding, Dihedral angle. All these forces will give more accurate docking energy.

**Authors**

**Uttam K. Roy** is presently Assistant Professor in the Department of Information Technology, Jadavpur University, Kolkata. He completed his M. Tech in Computer Science and Engineering, and PhD from Jadavpur University, Kolkata. For excellence in academics, he was awarded scholarships from UGC and Jadavpur University. In addition to his 12-year teaching experience, he has been a technical consultant and system administrator.

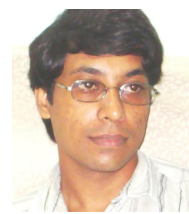

Dr Roy's research interests include bio-informatics, voice processing, optimization, and quantum computing. He is the sole author of two text books "Web Technologies" and "Advanced Java Programming", published by Oxford University Press, India in 2010. He also has contributed numerous research papers to various international journals, and has guided and supervised many postgraduate and Ph D dissertations.

**Mainak Talukdar** is presently working as a Programmer Analyst Trainee at Cognizant Technology Solutions, Kolkata. He has done Master of Engineering in Software Engineering from Jadavpur University. Earlier, he has done his Bachelor of Technology in Computer Science and Engineering from West Bengal University of Technology, Kolkata. He has about two year of industry experience.

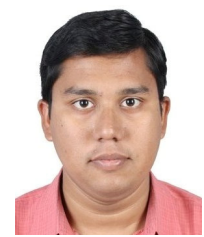

Mainak's research interests include bio-informatics, data warehousing.

**Anupam Ghosh** is presently working as a Research Engineer at the Center for Indian Language Technology, Department of Computer Science and Engineering, Indian Institute of Technology Bombay, Mumbai. He completed his Master of Engineering from Jadavpur University, Kolkata. Earlier, He has done his Bachelor of Technology in Computer Science and Engineering from West Bengal University of Technology, Kolkata. He has Academic experience of about 2 years and Industry experience of about 2 year.

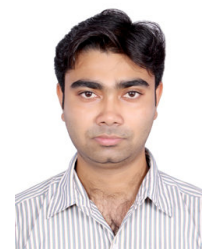

Anupam's research interests include bio-informatics, Natural Language Processing. He wishes to pursue his career in the field of Natural Language Process and Machine Leaning.